# In situ modification of delafossite-type $PdCoO_2$ bulk single crystal for reversible hydrogen sorption and fast hydrogen evolution

*Guowei Li,[Δ,†"] Seunghyun Khim,[Δ"] Celesta S. Chang,[^"] Chenguang Fu,[Δ] Nabhanila Nandi,[Δ] Fan Li,[§] Qun Yang,[Δ] Graeme R. Blake,[□] Stuart Parkin,[§] Gudrun Auffermann,[Δ] Yan Sun,[Δ] David A. Muller,[+¶] Andrew P. Mackenzie,[Δ±] and Claudia Felser[Δ,†]*

[Δ] Max Planck Institute for Chemical Physics of Solids, 01187 Dresden, Germany

[^] Department of Physics, Cornell University, Ithaca, New York, USA

[§] Max Planck Institute for Microstructure Physics, 06120 Halle, Germany

[□] Zernike Institute for Advanced Materials, University of Groningen, 9747 AG, Groningen, The Netherlands

[+] School of Applied and Engineering Physics, Cornell University, Ithaca NY 14853, USA

[¶] Kavli Institute at Cornell for Nanoscale Science, Cornell University, Ithaca, New York, USA

[±] Scottish Universities Physics Alliance, School of Physics and Astronomy, University of St. Andrews, Fife KY16 9SS, United Kingdom

ABSTRACT: The observation of extraordinarily high conductivity in delafossite-type $PdCoO_2$ is of great current interest, and there is some evidence that electrons behave like a fluid when flowing in bulk crystals of $PdCoO_2$. Thus, this material is an ideal platform for the study of the electron transfer processes in heterogeneous reactions. Here, we report the use of bulk single crystal $PdCoO_2$ as a promising electrocatalyst for hydrogen evolution reactions (HERs). An overpotential of only 31 mV results in a current density of $10\ \mathrm{mA\ cm^{-2}}$, accompanied by high long-term stability. We have precisely determined that the crystal surface structure is modified after electrochemical activation with the formation of strained Pd nanoclusters in the surface layer. These nanoclusters exhibit reversible hydrogen sorption and desorption, creating more active sites for hydrogen access. The bulk $PdCoO_2$ single crystal with ultra-high conductivity, which acts as a natural substrate for the Pd nanoclusters, provides a high-speed channel for electron transfer.

Delafossite-type metals such as $PdCoO_2$ are currently attracting considerable attention due to their exotic and surprising physical properties. The most striking characteristic is their remarkably low electrical resistivity, lower than those of all normal state oxides yet reported.[1-3] Pd 4$d$ states contribute to the density of states at the Fermi level with $sp$ band-like fast dispersion.[4-5] This leads to an ultra-high carrier velocity of ~ $4.96 \times 10^5$ m/s, which even approaches that of graphene with a velocity of the order of ~ $10^6$ m/s.[1, 6-8] Of particular interest is the observation of hydrodynamic electron flow in bulk $PdCoO_2$ single crystals, suggesting the anomalous suppression of momentum relaxation resulting from electron-impurity, electron-electron, and electron-phonon.[9] In addition, $PdCoO_2$ exhibits a strong quasi-two-dimensional nature, characterized by higher in-plane conductivity than along the normal direction. This represents an ideal research prototype to study the relationship between conductivity and surface reactions, especially in the field of electrochemistry.

Hydrogen production from electrochemical water splitting does not only have potential to face the challenge of global environmental and energy problems but also serves as a model for the study of other complex heterogeneous catalysis reactions.[10-15] It is commonly agreed that an ideal electrocatalyst for HER has a large exchange current density, a low Tafel slope, and high electrochemical stability. The efficiency of a given catalyst is mainly determined by the first two factors, which in turn, represent electron transfer barriers. Further investigation of the HER process revealed that two electron transfer processes are involved. The first one involves electron transfer from the catalyst surface to the reaction intermediate. This is governed by the Sabatier principle based on the d-band model, which states that an optimal binding energy of the intermediate will favor the reaction. The second process is the injection of electrons from the substrate (e.g. glassy carbon electrode) to the catalyst and then flows to the surface.

We can now ask two interesting questions. What is the HER performance of a bulk single crystal with strikingly low room-temperature resistivity? Can we combine the characteristics of high current density, low Tafel slope and high electrochemical stability in a single compound?

Herein, we report the synthesis of high-quality $PdCoO_2$ single crystals such that we can observe their high-conductivity (001) crystal surface. Serving as both an electrode and catalyst for HER, $PdCoO_2$ exhibits remarkable electrocatalytic activity with an extremely low overpotential, low Tafel slope, and good stability, outperforming the state-of-the-art Pt catalyst. Scanning transmission electron microscopy (STEM) revealed the nature of the precise surface modification process in the catalytic environment. The formation of a surface layer consisting of Pd nanoclusters and cobalt oxide is responsible for the high activity with respect to HER. We observe reversible hydrogen sorption and desorption in these Pd nanoclusters, which formed in-situ on the bulk $PdCoO_2$ single-crystal surface. This provides a large number of active sites for hydrogen access. The bulk single crystal substrate also served as an expressway for electron transfer. Our findings suggest that the in-situ modification of a high-conductivity bulk single crystal is crucial for tuning its electrocatalytic activity.

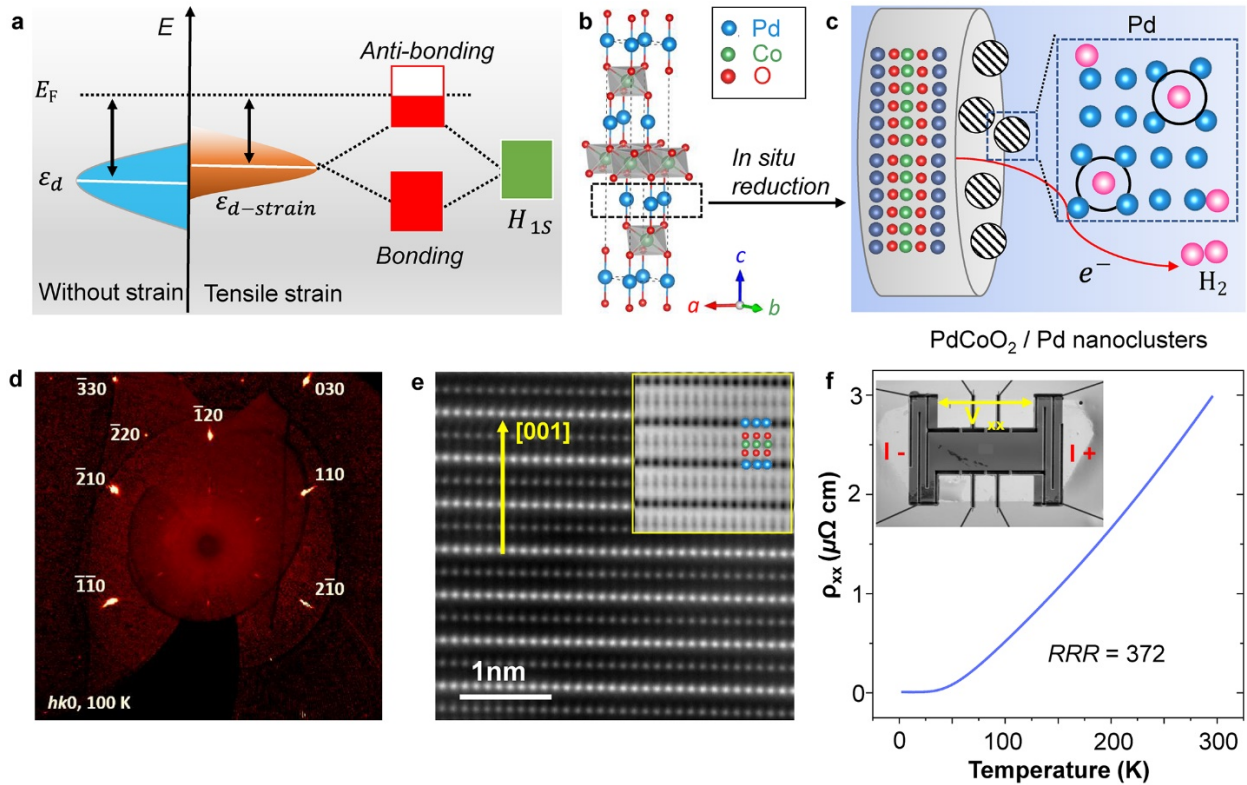

**Figure 1. a.** Changes in d-bands at the transition metal surface in the presence of tensile strain. A shift of the d-band center is expected when coupled with H adsorbates. **b.** Crystal structure of $PdCoO_2$ single crystal obtained by single crystal XRD. **c.** Hydrogen adsorption/absorption process and electron transfer process on the in-situ modified $PdCoO_2$ single crystal surface. Pink spheres are hydrogen atoms, and blue spheres are Pd atoms. **d.** Single-crystal XRD precession image of the *hk*0 reciprocal space plane of a bulk $PdCoO_2$ crystal. **e.** HADDF-STEM image showing the cross-section of $PdCoO_2$ single crystal [120] prepared by FIB. The inset shows an ABF-STEM image, resolving the oxygens clearly. The crystal structure is overlaid, where Pd, Co, and O atoms are represented by blue, green, and red circles respectively. **f.** Temperature-dependent electrical resistivity of the single crystal measured using a device fabricated by FIB as shown in the inset.

Let us first summarize our strategy and its advantages. For HER catalysts, it has been confirmed that the surface reactivity trends can be rationalized by the d-band theory (Figure 1a). The chemisorption of H on transition metals results in coupling between the H 1s band and the metal d-bands. The average energy of the electronic d-states projected onto a surface metal atom (the d-band center $\varepsilon_d$) determines the filling of adsorbate-metal antibonding states and further controls the activity. However, once strain is introduced in the transition metal, the interatomic bond lengths and the environment can change. For transition metals with more than half filled d-bands, a tensile strain should upshift the d-band center to preserve the degree of d-band filling.[16-17] Generally, strain can be introduced either by preparing nanostructures or by depositing on the substrate with mismatched crystal lattices, where one can observe the strain at the interface, within the facets, or at the edges of the catalyst. Unfortunately, these strategies generally result in an uncertain distribution and extent of strain. $PdCoO_2$ adopts a quasi-two-dimensional structure, with Co atoms forming distorted octahedra with oxygen atoms, and are interlinked by single-Pd layers along the $c$ axis.[2] As a cathode material for HER, we expect the in-situ formation of Pd nanoclusters during the reduction process (Figure 1b). Such a layer-by-layer growth process at the expense of $PdCoO_2$ may introduce strain not only confined to the crystal surface, but also in the bulk. The formation of Pd nanoclusters is extremely important for hydrogen adsorption and absorption in the subsequent hydrogen reduction process (Figure 1c).

Single crystals of $PdCoO_2$ were grown via a reaction between stoichiometric amounts of $PdCl_2$ and CoO.[9, 18] A typical single crystal is displayed in Figure S1, showing metallic luster and well-defined steps typical of a layered structure. The atomic stoichiometries of Co, Pd, and O are determined to be approximately 1:1:2 for both the surface and the sides (Figure S2), indicating high homogeneity of the bulk single crystal. More accurately, de Haas-van Alphen measurements of the Fermi volume confirm stoichiometry to within the ~ 1% accuracy of that technique.[8] The high purity of the sample is confirmed by the X-ray diffraction (XRD) pattern recorded on the powder crushed from bulk single crystal (Figure S3). All the diffraction peaks can be indexed to PdCoO$_2$ with space group $R\bar{3}m$ (space group no. 166). More detail are given by single-crystal X-ray diffraction at 100 K and 295 K (Figure 1b). PdCoO$_2$ crystallizes in the hexagonal system with space group R-3m and lattice parameters of $a = b = 0.2845(3)$ nm and $c = 1.7880(14)$ nm at 295 K. The crystal is highly layered leading to some smearing and splitting of spots in the stacking direction (see the 0$kl$ reciprocal lattice plane in Figure S4), probably caused by slight rotations of the layers along the stacking direction. In contrast, the degree of in-plane order is very good, as can be seen by the sharp spots in the $hk$0 reciprocal lattice plane (Figure 1d). A cross-sectional TEM lamella was prepared from the as-grown single crystal using a focused ion beam (FIB) (Figure S5). The lattice structure clearly shown in the enlarged STEM image shows the high crystallinity of the PdCoO$_2$ crystal (Figure 1e). High-angle annular dark-field (HAADF-STEM) image emphasizes the repeating layers of Pd and CoO; the inset annular bright-field (ABF-STEM) image also shows the oxygen atoms in the CoO layer. HAADF-STEM image that looking down on [001] crystal zone proved the exposing of (001) crystal surface (Figure S6). The surface composition and electronic structure were investigated by X-ray photoelectron spectroscopy (Figure S7). The prominent Pd 3d$_{5/2}$ peak at 335.4 eV is assigned to the Pd-O bonds in PdCoO$_2$ (Figure S8). This binding energy is slightly larger than that of pure Pd (334.8- 335.1 eV), but significantly lower than in PdO.[19-20] This indicates that Pd adopts the monovalent state in the PdCoO$_2$ crystal. Thus, the Pd-O layer should be a good metal from the viewpoint of the ionic picture, which is further confirmed by the asymmetric shape of the peak and the plasmonic energy loss feature.[7] Extensive photoemission studies of surface states have quantified the level of charge redistribution at Co-O terminated [21] and Pd-terminated [22] surfaces of bulk crystals.

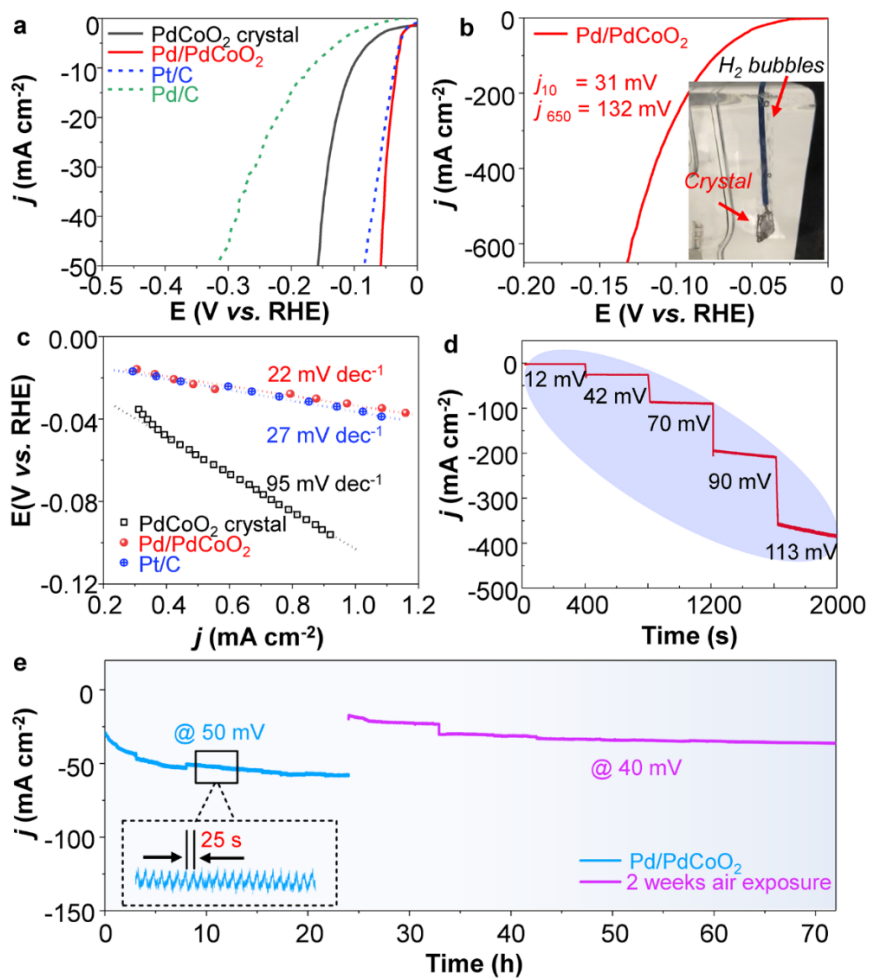

**Figure 2. a.** Polarization curves of the modified bulk PdCoO$_2$ single crystal and 20 % Pt/C, Pd/C commercial catalysts. **b.** Polarization curves of the bulk PdCoO$_2$ single crystal at larger applied overpotential. The inset shows the produced hydrogen bubbles on the crystal surface. **c.** Tafel plots of the bulk PdCoO$_2$ single crystal catalysts and 20 % Pt/C catalyst. **d.** Multi-current process showing current density increasing from 5 to 350 mA cm$^{-2}$ without iR correction. **e.** Current versus time over 24 h showing a constant potential of -50 mV, which is reduced to -40 mV (for 48 h) after two weeks of exposure to air.

In this study, the same PdCoO$_2$ bulk single crystals were used as both the cathode and electrocatalyst. Thus, their electrical conductivity is extremely important for the HER activity. Devices with well-defined geometry were fabricated by FIB sculpting as shown in the inset of Figure 1f. The temperature dependence of the resistivity exhibits the behavior of a metal. The room-temperature resistivity was only 2.9 µΩ cm, one of the lowest electrical resistivities among all reported metal oxides. This value is even lower than those of benchmark metal catalysts such as Pt (10.6 µΩ cm), Pd (10.5 µΩ cm), Ru (7.1 µΩ cm), and Ir (4.7 µΩ cm). The residual resistivity ratio (RRR), R(300K) / R(2K), is 372, indicating the high quality of the bulk crystal.

The well-defined exposed crystal surface, as well as the ultra-high room temperature conductivity, makes the crystal suitable for studying the surface HER mechanism. The electrochemical characteristics of the bulk single crystal were measured using a standard three-electrode setup in 0.5 M H$_2$SO$_4$ (aq). The *iR* corrected polarization curves of PdCoO$_2$ and a 20 % Pt/C catalyst are shown in Figure 2a. To achieve a catalytic current density of 10 mA cm$^{-2}$, the modified PdCoO$_2$ single crystal requires an overpotential of 31 mV, which is even lower than that of commercial Pt/C (35 mV). More strikingly, a very small overpotential of only 132 mV was needed to deliver a current density of 650 mA cm$^{-2}$ (Figure 2b). Hydrogen bubbles are produced furiously on the crustal surface (Inset Figure 2b). To the best of our knowledge, this represents one of the best electrocatalysts with such high HER activity (Table S1).[23-32] Tafel slopes of 22 and 27 mV dec$^{-1}$ were obtained for the modifed PdCoO$_2$ single crystal and Pt/C catalyst from the Tafel plots (Figure 2c). This suggests that the Tafel recombination step is the rate-determining step and is responsible for the particularly high activity of single crystal PdCoO$_2$, similar to the benchmark Pt catalyst, [33-36] The rapid response of the current density to voltage, as obtained from the multiple-step chrono-potentiometry experiments, revealed the high mass transport and mechanical robustness of the single crystal (Figure 2d).

The single crystal catalyst also exhibited outstanding long-term operation stability over 24 h at a constant overpotential (Figure 2e). By magnifying the catalytic current density versus time plot, a wave mode signal was observed, originating from the rapid hydrogen bubble release (inset to Figure 2e). Remarkably, after exposing the same crystal to air for 2 weeks, it still shows high stability beyond 48 h. Finally, the amount of the $H_2$ gas produced was collected at an HER current of 4 mA. A comparison with the theoretical value revealed a Faradaic efficiency of ~ 100 % (Figure S9).

Next, we will attempt to understand the origin of the high HER activity of the $PdCoO_2$ single crystal. The experiment was repeated using two fresh crystals, showing similar behavior as discussed below. Specifically, the fresh crystals did not show such impressive activity in the initial test (Figure 2a). A much larger overpotential of 105 mV was needed to achieve a current density of 10 mA cm$^{-2}$, accompanied by a steeper Tafel slope of 95 mV dec$^{-1}$ (Figure 2a and c). This may explain why delafossite $PdCoO_2$ has not received much attention as an HER catalyst. The activation process can be further elucidated by the stability test (Figure 2e). At an overpotential of ~50 mV, the catalytic current density increases continuously with time. Thus, we conclude that the surface of fresh single crystal $PdCoO_2$ must be re-constructed and optimized to favor HER kinetics.

A SEM image of a fresh crystal reveals a smooth and flat surface (Figure 3a). In contrast, many coniform islands are observed after HER measurement (Figure 3b). To elucidate the detailed structure of these islands, a cross-sectional TEM lamella, showing these features, was fabricated using a FIB for TEM analysis (Figure 3c and Figure S10-11). A comparison of the TEM images of a fresh crystal (Figure 3f) and an activated crystal (Figure 3d) indicate that in the activated crystal, a new surface layer of up to tens of nanometres is observed, as pointed out by the yellow arrow. Magnified STEM images of this area are shown in Figure 3e and g, where numerous polycrystalline Pd nanoparticles and cobalt oxide inclusions are observed in the new layer with distinctly different lattice structures from bulk $PdCoO_2$. Electron energy loss (EELS) chemical mapping was performed carefully across the interface (Figure 3g and Figure S12). The bright spots in the oxygen map overlap exactly with the Co signal in the cobalt map, whereas there is no Pd signal corresponding to these regions. The elemental color mapping further confirms that the newly formed surface layer contains large polycrystalline Pd clusters at the top, followed by an intermediate layer comprising small Pd nanoclusters and cobalt oxide inclusions, with the $PdCoO_2$ bulk crystal at the bottom showing clearly distinguishable periodic Pd and CoO layers. From the multivariate curve resolution (MCR) processing of the resolved EELS spectra, we observe a slight shift of the Co spectrum peak from the nanoparticle layer with respect to that from the bulk $PdCoO_2$. We determined that the cobalt in the surface layer is in the $Co^{2+}$ valence state by comparing the obtained cobalt $L_{2,3}$-edge EELS spectrum with a reference spectrum (Figure S13).

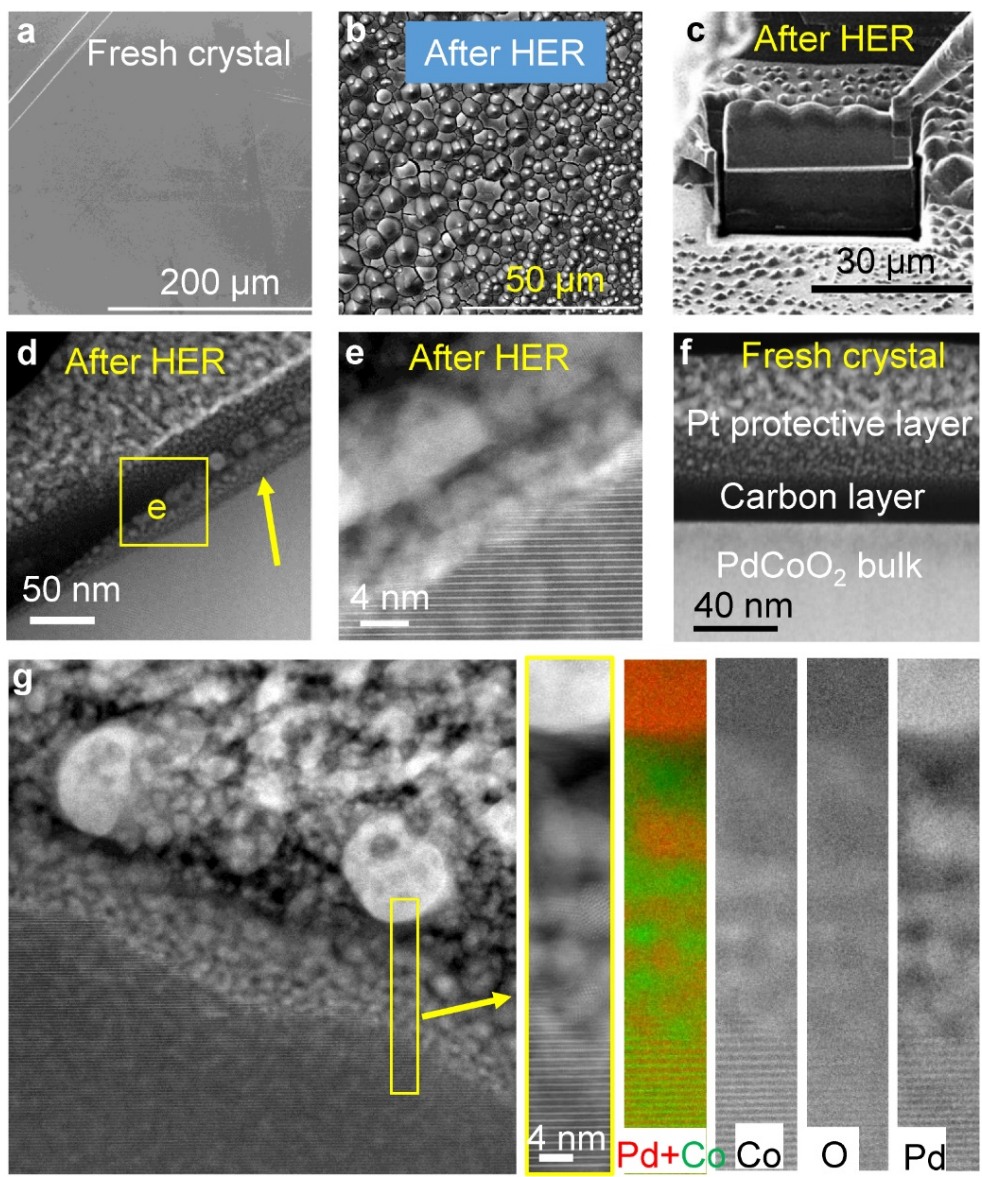

**Figure 3. a.** SEM image of a fresh PdCoO$_2$ single crystal surface. **b.** SEM image of the crystal surface after HER. **c.** A thin lamella including three islands is cut out by FIB. **d.** HAADF-STEM image of the single crystal surface after HER. A new surface layer is observed as indicated by the yellow arrow. **e.** Magnified STEM image of the interface indicated by the square in **d**. **f.** HAADF-STEM image of a fresh single crystal surface before HER. Compared to **d** and **e**, there are no surface layers. **g.** EELS elemental mapping across the interface. Pd and Co are clearly separated. O is detected only in the area where Co is present. We thus conclude that Pd nanoclusters and cobalt oxides are formed in the surface layer.

We studied the HER activities of various reported Pd and CoO structures (Table S2), and concluded that activity as impressive as that obtained herein as never been achieved before, especially with CoO phase catalysts. This suggests that the in-situ formation of a surface layer comprising Pd nanoclusters is closely related to the remarkable HER activity of bulk $PdCoO_2$ crystals. Cyclic voltammetry (CV) was next used to understand the hydrogen adsorption and absorption behavior on the crystal surface. Figure 4a shows that the cathodic and anodic peaks between 0.25 and 0.4 V corresponds to hydrogen adsorption and formation of the α-phase (Pd +

H) at the Pd nanocrystal surface. The formation of the β-phase hydride (PdH$_x$) is observed at a low potential and saturates at ~ 0 mV. This is followed by the hydrogen evolution process when the potential *vs.* RHE is negative. These features are in good agreement with previous results obtained using various Pd structures.[37] To achieve full saturation of hydrogen at the crystal surface, a much slower scan rate of 1 mV/s was used. The corresponding CV curves (Figure 4b) provides two important results. First, the hydrogen adsorption/desorption and absorption/desorption charges are almost the same during the potential scans. Second, the potential hysteresis between the adsorption and absorption process is only 24 mV and 14 mV, significantly lower than that obtained for Pd nanoparticles of the same size and other Pd structures. Since the hysteresis reflects electrochemical reversibility of the hydrogen absorption/desorption processes, this implies high reversibility in the present system.[38]

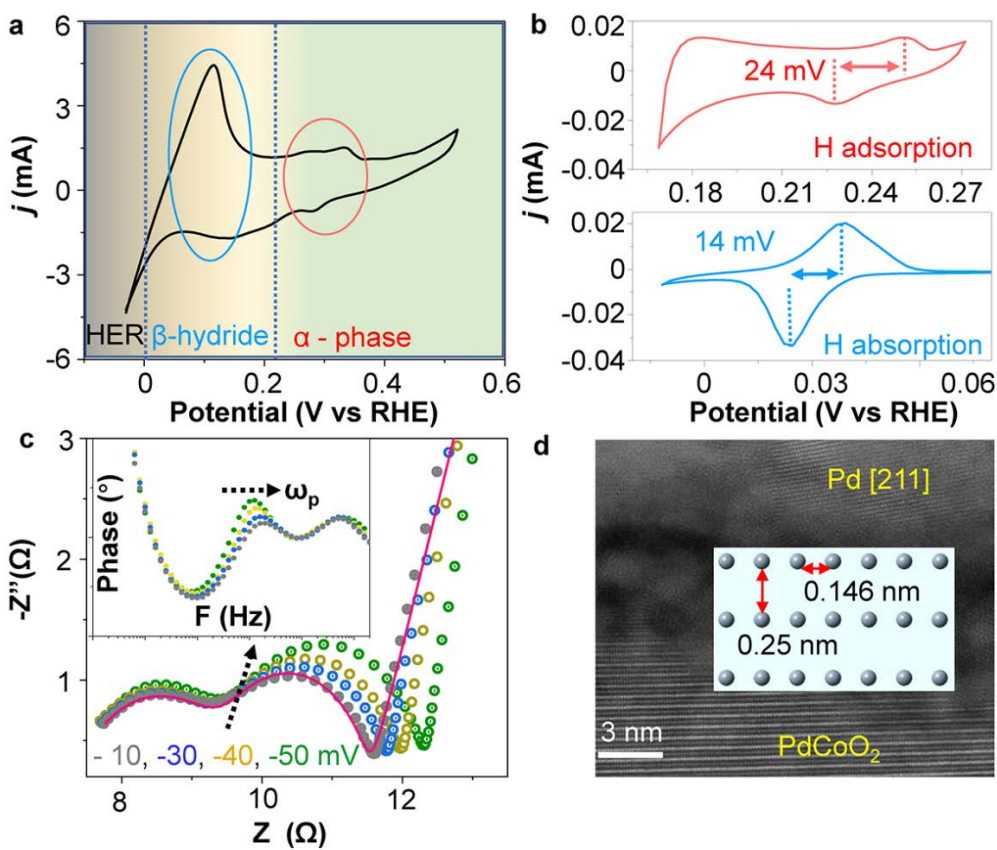

**Figure 4. a.** CV curve of PdCoO$_2$ single crystal at a scan rate of 20 mV / s. The formation of various phases in different potential ranges are separated by dotted lines. **b.** CV curves recorded in the hydrogen adsorption and absorption range at a scan rate of 1 mV / s, corresponding to the formation of the α-phase and β-phase, respectively. **c.** Nyquist plots of the PdCoO$_2$ single crystal with different overpotentials. The corresponding Bode phase plots are shown in the inset. **d.** HAADF-STEM image of the surface modified PdCoO$_2$ single crystal. The measured apparent spacings between Pd atoms on the (211) projection of a Pd nanocluster are 0.25 and 0.146 nm, which are significantly larger than the corresponding distances for pure Pd.

Electrochemical impedance spectroscopy (EIS) was used to investigate the electron transfer and hydrogen sorption mechanisms. The EIS patterns at different potentials were recorded in the overpotential deposition (OPD) region, corresponding to the HER process (Figure 4c). The patterns can be fitted by a two-time constant serial (2TS) model, with two semicircles observed in the Nyquist plots (Figure S14). The semicircle at high frequency (HF), which is strongly independent of the overpotential, can be assigned to electron transfer from the electrode to the catalyst.[39-40] More specifically, this corresponds to electron transfer from the bulk PdCoO$_2$ single crystal to the surface Pd nanoparticles. The low frequency (LF) semicircle depends on the overpotentials. This is related to the charge transfer kinetics in the HER process, corresponding to a faster reaction rate. Bode phase diagram reveals two relaxation times (Figure 4c and Figure S15). With increasing overpotential, the intensity of the LF semicircle decreases and shifts to a higher frequency, indicating an increase in the reaction rate and a shorter reaction time constant ($\tau = 1/\omega_p$, where $\tau$ is the time constant and $\omega_p$ is the characteristic frequency).[41-42] In contrast, the time constant at HF has a slight decrease in density and is much lower than the HER kinetics value. This corresponds to the response of a fast electron transfer process, as proposed by Omanovic et al.[43] Thus, we can conclude that the high electrical conductivity of the PdCoO$_2$ single crystal is of great importance for electron injection onto the catalyst surface.

The formation of the β-phase PdH$_x$ as a result of hydrogen sorption, also leads to another phenomenon. Figure 4d shows that the average apparent distances between Pd atoms in the (211) plane projection of the Pd nanoclusters are 0.25 nm and 0.16 nm, which are significantly larger than the corresponding distances in pure Pd (0.25 nm and 0.137 nm) (more details in Figure S16). This corresponds to anisotropic tensile strains of 10 % and 6.2 % along the longer and shorter length of Pd, respectively. The volcano plot obtained based on the Sabatier principle suggests that the adsorption energy of reaction intermediates are always very high on transition metal surfaces, leading to sluggish HER kinetics.[44-45] The introduction of strain on the catalyst surface has been proven as an effective strategy to regulate the adsorbate energies.[46] A shift of the d-band center was observed for various transition metals including Pd, Pt, Ir, Ni, and Co, when the strain is created, regardless of the exposed catalyst surface.[47] Indeed, our calculations suggests an up-shifted of the *d* band center from -2.30 eV for the pristine sample to -2.19 eV after applying a tensile strain (Figure S17). An attractive point of our current study is, rather than following strategies like creating defects, alloying and depositing on a substrate with different lattice constants, that strain is introduced in-situ during the HER process on the highly conducive PdCoO$_2$ single crystal surface.[44, 48-49] Problems like mechanical instability, strain relaxation, discontinuous strain distribution, and dissolution can be avoided by this method.

In summary, high-quality bulk single crystals of PdCoO$_2$ have been synthesized and its electrochemical HER behavior has been studied systematically. We observe reconstruction of the crystal surface, leading to the in-situ formation of a surface layer comprising Pd nanoclusters and cobalt oxide. It is demonstrated that these Pd nanoclusters exhibit high hydrogen sorption/desorption reversibility, which creates more active sites for hydrogen to access during the subsequent HER process. In addition, the ultra-high conductivity of the PdCoO$_2$ single crystal substrate provides an expressway for electron injection into the catalyst surface. The methodology presented herein provides a general way for in-situ modification and fabrication of high-efficiency HER electrocatalysts.

## AUTHOR INFORMATION

G. L, S. K. C. S. C. contributed equally to this work. Corresponding authors: Guowei.li@cpfs.mpg.de; Claudia.felser@cpfs.mpg.de

## ACKNOWLEDGMENT

This work was financially supported by the European Research Council (ERC Advanced Grant No. 291472 'Idea Heusler') and ERC Advanced Grant (No. 742068). TOPMAT'. Electron Microscopy (C.C., D.M.) supported by the U.S. Department of Energy, Basic Energy Sciences grant No. DE-SC0019445, with facility support from the National Science Foundation NSF Grant No. DMR-1719875. Raw data for all figures in this paper are available at (URL to be provided).